\documentstyle[multicol,aps,psfig]{revtex}
\renewcommand{\narrowtext}{\begin{multicols}{2}
\global\columnwidth20.5pc\noindent}
\renewcommand{\widetext}{\end{multicols}
\global\columnwidth42.5pc}
\begin{document}
\draft
\preprint{12 November 2003}
\title
{Static and Dynamic Properties of Antiferromagnetic Heisenberg Ladders:\\
 Fermionic versus Bosonic Approaches}
\author{Hiromitsu Hori and Shoji Yamamoto}
\address{Division of Physics, Hokkaido University,
         Sapporo 060-0810, Japan}
%\date{Received \hspace{4cm}}
\date{Received 12 November 2003}
\maketitle
\begin{abstract}
In terms of spinless fermions via the Jordan-Wigner transformation along
a snake-like path and spin waves modified so as to restore the sublattice
symmetry, we investigate static and dynamic properties of two-leg
antiferromagnetic Heisenberg ladders.
The specific heat is finely reproduced by the spinless fermions, whereas
the magnetic susceptibility is well described by the modified spin
waves.
The nuclear spin-lattice relaxation rate is discussed in detail with
particular emphasis on its novel field dependence.
\end{abstract}
% 75.10.Jm: Quantized spin models
% 75.30.Cr: Saturation moments and magnetic susceptibilities 
% 75.30.Ds: Spin waves (for spin-wave resonance, see 76.50.+g)
% 75.40.Cx: Static properties (order parameter, static susceptibility,
%           heat capacities, critical exponents, etc.) 
% 75.40.Gb: Dynamic properties (dynamic susceptibility, spin waves,
%           spin diffusion, dynamic scaling, etc.)
% 75.40.Mg: Numerical simulation studies
% 75.50.Xx: Molecular magnets
% 76.50.$+$g: Ferromagnetic, antiferromagnetic, and ferrimagnetic
%             resonances; spin-wave resonance
\narrowtext

   Spin gaps$-$the energy gaps in magnetic excitation spectra$-$have been
attracting much interest in recent years.
Haldane \cite{H464,H1153} pioneeringly pointed out possible gapped
excitations in one-dimensional Heisenberg antiferromagnets with integral
spins.
There followed interesting topics such as spin gaps in a magnetic field
\cite{O1984} and mass generation in mixed-spin chains \cite{F14709,F8799}.
In such circumstances, Dagotto {\it et al.} \cite{D5744} stimulated new
interest in Heisenberg ladder antiferromagnets suggesting another
electronic mechanism for the gap formation.

   There are a variety of analytic, as well as numerical, studies on
ladder antiferromagnets.
The strong-coupling expansion from decoupled dimers \cite{B3196} and the
bond-operator representation of rung dimers \cite{G8901} are the very
methods to treat two-leg ladders and indeed clarified the nature of
low-lying excitations.
However, they are less applicable to thermodynamics.
Although resonating valence-bond wave functions \cite{L365} may be still
relevant to ladder materials, their variational calculation, even on a
nearest-neighbor singlet basis \cite{F1820,B333}, is practically
restricted to ground-state properties \cite{W886,Z11609}.
Then we may consider making the best use of conventional tools such as
spin waves and Jordan-Wigner spinless fermions.

   The naivest generalization of the Jordan-Wigner transformation to
coupled spin chains \cite{S2419} enlighteningly visualizes the gap
formation with existing interchain interaction, but the obtained spin gap
\cite{A6131} is less precise in the intermediate interchain-coupling
regime of our interest.
Recently, performing the Jordan-Wigner transformation along an elaborately
ordered path but simplifying the mean-field treatment, several authors
\cite{D964,H1607} obtained a better description of the spin gap for an
arbitrary number of legs.
We develop their scheme to thermal and dynamic properties.
Another interest in this article is to examine a spin-wave picture for
spin-gapped antiferromagnets \cite{Y769}.
The conventional spin-wave theory \cite{A694,K568} applied to Heisenberg
single-rung ladders ends in diverging sublattice magnetizations and
vanishing gap \cite{B3196}.
We therefore employ the sublattice-symmetric spin-wave theory
\cite{T2494,H4769}, originally applied to square-lattice antiferromagnets.
Handling Holstein-Primakoff bosons, we make our first attempt to describe
ladder antiferromagnets in terms of spin waves.
{\it We make interesting explorations into ladder systems featuring
fermions versus bosons} in an attempt to provide an intuitive picture for
understanding the numerical findings and to reveal novel quantum phenomena
peculiar to low dimensions.

   We study the spin-$\frac{1}{2}$ antiferromagnetic Heisenberg model on
two-leg ladders:
\begin{equation}
   {\cal H}
   =J\sum_{i=1}^2\sum_{j=1}^N
    \mbox{\boldmath$S$}_{i,j}\cdot\mbox{\boldmath$S$}_{i,j+1}
   +J'\sum_{j=1}^N
    \mbox{\boldmath$S$}_{1,j}\cdot\mbox{\boldmath$S$}_{2,j},
   \label{E:H}
\end{equation}
which has isotropic nearest-neighbor exchange interactions along the
chains ($J$) and along the rungs ($J'$).
In the two extreme cases of $J'=0$ and of $J=0$, the low-energy properties
are well known.
The decoupled chains are critical, whereas the decoupled rung dimers are
massive.
The intermediate region with $J'\simeq J$ is thus interesting, but most of
model materials lie in the regime of large $J'$, such as
Cu$_2$(C$_5$H$_{12}$N$_2$)$_2$Cl$_4$ with $J'\simeq 5J$ \cite{H5392},
Cu$_2$(C$_5$H$_{12}$N$_2$)$_2$Br$_4$ with $J'\simeq 7J$ \cite{D1599}, and
(C$_5$H$_{12}$N)$_2$CuBr$_4$ with $J'\simeq 3.5J$ \cite{W5168}.
Hence much attention is paid to a copper oxide SrCu$_2$O$_3$, comprising
two-leg ladders with $J'\simeq J$ \cite{A3463}.
We consider the case of $J'=J$ unless otherwise noted.

   It is along a snake-like path \cite{D964,H1607},
$(i,j)=(1,1)\rightarrow(2,1)\rightarrow
       (2,2)\rightarrow(1,2)\rightarrow
       (1,3)\rightarrow\cdots$,
that we align spinless fermions (SFs).
When we introduce renumbered spin operators
$\widetilde{\mbox{\boldmath$S$}}_{i,j}=\mbox{\boldmath$S$}_{i,j}$
($\mbox{\boldmath$S$}_{\bar{i},j}$) for odd (even) $j$'s, where
$\bar{i}=3-i$, the SFs are created as
$
   c_{i,j}^\dagger=\widetilde{S}_{i,j}^+
   {\rm exp}[-{\rm i}\pi
   (\sum_{n=1}^{j-1}\sum_{m=1}^2
     \widetilde{S}_{m,n}^+\widetilde{S}_{m,n}^-
   +\sum_{m=1}^{i-1}
     \widetilde{S}_{m,j}^+\widetilde{S}_{m,j}^-)]
$.
The fermionic Hamiltonian is given by
\begin{eqnarray}
   &&
   {\cal H}
   =\sum_{i=1}^2\sum_{j=1}^N
    \Bigl[
     J\bigl(
       c_{i,j}^\dagger c_{\bar{i},j+1}^\dagger c_{\bar{i},j+1} c_{i,j}
      -c_{i,j}^\dagger c_{i,j}+\frac{1}{4}
      \bigr)
   \nonumber\\
   &&\qquad
     +\frac{J}{2}
      \bigl(
       c_{i,j}^\dagger c_{\bar{i},j+1}
       {\rm e}^{-{\rm i}\pi\delta_{i1}
                (c_{1,j+1}^\dagger c_{1,j+1}+c_{2,j}^\dagger c_{2,j})}
      +{\rm H.c.}
      \bigr)
   \nonumber\\
   &&\qquad
   +\frac{J'}{2}
     \bigl(
      c_{i,j}^\dagger c_{\bar{i},j}^\dagger c_{\bar{i},j} c_{i,j}
     -c_{i,j}^\dagger c_{i,j}+c_{i,j}^\dagger c_{\bar{i},j}+\frac{1}{4}
     \bigr)
    \Bigr],
\end{eqnarray}
and we search for its mean-field solutions.
Setting the thermal average $\langle c_{i,j}^\dagger c_{i,j}\rangle$ equal
to $1/2$ under zero magnetization and defining a unitary transformation
\begin{equation}
   \left(
    \begin{array}{c}
     \alpha_{1,k}\\
     \alpha_{2,k}
    \end{array}
   \right)
  =\frac{1}{\sqrt{2}|\varepsilon_k|}
   \left(
    \begin{array}{cc}
     |\varepsilon_k| &
      \varepsilon_k  \\
      \varepsilon_k^*&
    -|\varepsilon_k|
    \end{array}
   \right)
   \left(
    \begin{array}{c}
     c_{1,k} \\
     c_{2,k}
    \end{array}
   \right),
\end{equation}
we can diagonalize the Hamiltonian as
$\sum_k |\varepsilon_k|
 (\alpha_{1,k}^\dagger \alpha_{1,k}-\alpha_{2,k}^\dagger \alpha_{2,k})$,
where
\begin{eqnarray}
   &&
   \varepsilon_k
   =\Bigl(
     \frac{1}{2}-\chi_0
    \Bigr)J'
   +\Bigl(
     \frac{1}{2}-\chi_1-2|\chi_1|^2-\chi_2^*
    \Bigr)J\cos k
   \nonumber\\
   && \qquad
   +{\rm i}
    \Bigl(
     \frac{1}{2}-\chi_1+2|\chi_1|^2+\chi_2^*
    \Bigr)J\sin k,
    \label{E:dspSF}
\end{eqnarray}
with
$\chi_0=\langle c_{2,j}^\dagger c_{1,j}\rangle$,
$\chi_1=\langle c_{2,j}^\dagger c_{1,j+1}\rangle$, and
$\chi_2=\langle c_{1,j}^\dagger c_{2,j+1}\rangle$ to be self-consistently
determined at each temperature in an approximation of the Hartree-Fock
type, while within the Hartree-level approximation,
$\varepsilon_k=J'/2+{\rm i}J\sin k$, suggesting a spin gap $J'/2$.

   A bosonic approach starts from the Holstein-Primakoff transformation:
$S_{i,j}^+=(2S-a_{1,j}^\dagger a_{1,j})^{1/2}a_{1,j}$,
$S_{i,j}^z=S-a_{1,j}^\dagger a_{1,j}$ for even $i+j$'s, while
$S_{i,j}^+=a_{2,j}^\dagger(2S-a_{2,j}^\dagger a_{2,j})^{1/2}$,
$S_{i,j}^z=-S+a_{2,j}^\dagger a_{2,j}$ for odd $i+j$'s.
Then the Hamiltonian is expanded as
${\cal H}=\sum_{i=-2}^\infty{\cal H}_i$, where ${\cal H}_i$ is the
$O(1/S^i)$ term and is taken into consideration up to $i=0$.
${\cal H}_{-1}$ and ${\cal H}_0$ contain bilinear and biquadratic terms
with respect to the bosonic operators and describe the linear and
interacting spin waves, respectively \cite{Y14008}.
In order to avoid quantum as well as thermal divergence of the sublattice
magnetizations, the Bogoliubov transformation is carried out subject to
the constraint that the sublattice magnetizations be zero
\cite{T2494,H4769}:
\begin{figure}
\centerline
{\mbox{\psfig{figure=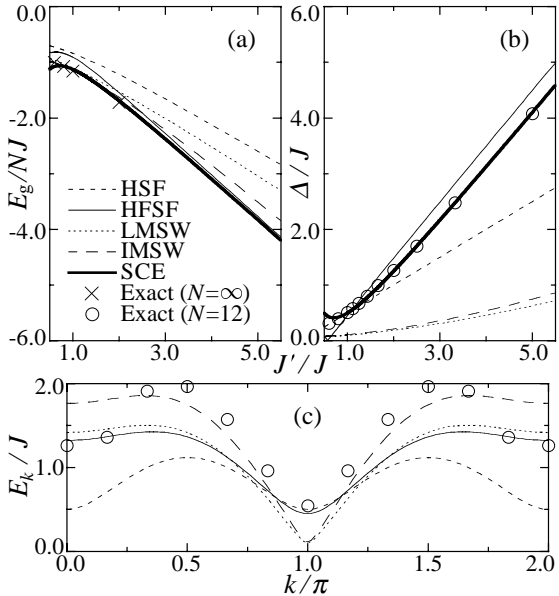,width=74mm,angle=0}}}
\vspace*{1mm}
\caption{The ground-state energy (a) and the spin gap (b) as functions of
         $J'/J$, and the dispersion relation of spin-triplet excitations
         (c).}
\label{F:GS}
\end{figure}
\begin{equation}
   \sum_j a_{1,j}^\dagger a_{1,j}
  =\sum_j a_{2,j}^\dagger a_{2,j}
  =NS.
  \label{E:const}
\end{equation}
Compared with the conventional spin-wave theory, where spins on one
sublattice point predominantly up, while those on the other predominantly
down, the modified spin waves (MSWs) restore the sublattice symmetry.
If we define the Bogoliubov transformation as
\begin{equation}
   \left(
    \begin{array}{c}
     a_{1,k}\\
     a_{2,k}^\dagger
    \end{array}
   \right)
  =\frac{1}{\sqrt{2}}
   \left(
    \begin{array}{cc}
     x_{1,k} & x_{2,k} \\
     x_{2,k} & x_{1,k}
    \end{array}
   \right)
   \left(
    \begin{array}{c}
     \beta_{1,k} \\
     \beta_{2,k}^\dagger
    \end{array}
   \right),
\end{equation}
${\cal H}_{-1}$ is diagonalized with
\begin{equation}
   \begin{array}{ccl}
    x_{1,k} &=& 
    \sqrt{\frac{(J'+2J)S+2J\lambda}{\varepsilon_k}+1},\\
    x_{2,k} &=& 
    \sqrt{\frac{(J'+2J)S+2J\lambda}{\varepsilon_k}-1}\,
    {\rm sgn}(J'-2J\cos k),
   \end{array}
\end{equation}
where $\lambda$ is the Lagrange multiplier due to the constraint
(\ref{E:const}) and $\varepsilon_k$ is the MSW dispersion relation:
\begin{equation}
   \varepsilon_k
   =\sqrt{[(2J+J')S+2J\lambda]^2-[(J'-2J\cos k)S]^2}.
   \label{E:dspMSW}
\end{equation}
The optimum thermal distribution functions of the MSWs are given by
$\langle\beta_{i,k}^\dagger\beta_{i,k}\rangle\equiv \bar{n}_{i,k}
 =[{\rm e}^{(\varepsilon_k-g\mu_{\rm B}H\cos i\pi)/k_{\rm B}T}-1]^{-1}$.
We consider ${\cal H}_0$ as a perturbation to ${\cal H}_{-1}$
\cite{N1380}.

   First we calculate the ground-state energy $E_{\rm g}$ and the spin gap
${\mit\Delta}$ by the linear MSWs (LMSWs), the interacting MSWs (IMSWs),
the Hartree-level SFs (HSFs), and the Hartree-Fock-level SFs (HFSFs), and
in Fig. \ref{F:GS} show them all together with numerical (Exact) findings
\cite{B3196} and strong-coupling-expansion (SCE) results \cite{R9235},
$E_{\rm g}/NJ=-3r/4-3/8r+O(r^{-2})$ and
${\mit\Delta}/J=r-1+1/2r+O(r^{-2})$, where $r=J'/J$.
Unless $J'$ is sufficiently large, the ground-state energy is better
described by the MSWs, while the spin gap by the SFs.
When we focus on the interesting point $J'=J$, the IMSWs give the best
estimate of the energy as $E_{\rm g}/NJ=-1.144$, compared with the
numerically exact value $-1.156$, whereas the HSFs give that of the gap as
${\mit\Delta}/J=1/2$, compared with $0.50(1)$.
The HSF description is no more precise once the system moves away from
this point, but the HFSF description remains quantitative in a wider range
of $J'/J$.
The MSWs are trivially less relevant to decoupled clusters, where no
``wave" can spread over the system.
The dispersion relation is also shown in Fig. \ref{F:GS}.
The band bottom is well reproduced by the SFs, while the band width by the
MSWs.
Since the MSWs considerably underestimate the spin gap, any thermally
activated behavior at low temperatures should be observed through the SFs.

\begin{figure}
\centerline
{\mbox{\psfig{figure=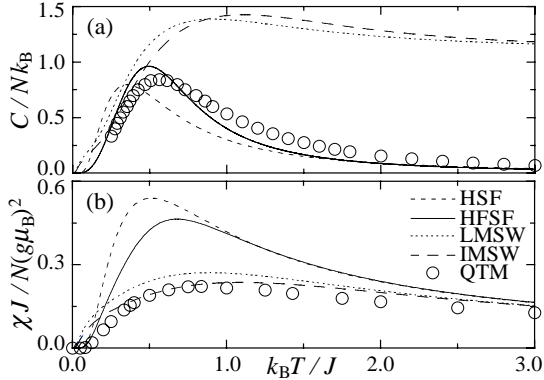,width=72mm,angle=0}}}
%\vspace*{1mm}
\caption{The specific heat (a) and the magnetic susceptibility (b) as
         functions of temperature.}
\label{F:Th}
\end{figure}

   Secondly we calculate the thermal properties and compare them with
quantum transfer-matrix (QTM) calculations \cite{T13515}.
The specific heat $C$ is shown in Fig. \ref{F:Th}(a).
The SFs well reproduce the Schottky peak, while the MSWs lose their
validity with increasing temperature.
The HSFs (HFSFs) underestimate the high-temperature behavior
$C/Nk_{\rm B}=(3k_{\rm B}T/4J)^2+O[(k_{\rm B}T/J)^4]$ by a factor $2/3$
($4/9$) but complement numerical tools very well at low temperatures.
If we approximate the dispersion relation of the low-lying excitations as
\begin{equation}
   E_k={\mit\Delta}+Ja(k-\pi)^2,
   \label{E:adsp}
\end{equation}
where ${\mit\Delta}=J'/2$ and $a=J/J'$ in the Hartree approximation, while
${\mit\Delta}=|(1/2-\chi_0)J'-(1/2-\chi_1-2\chi_1^2-\chi_2)J|$ and
$a=(1/2-\chi_0)
   [(1/2-\chi_1-2\chi_1^2-\chi_2)J'+4(\chi_2+2\chi_1^2)J]/2{\mit\Delta}$
in that of the Hartree-Fock type, the SFs illuminate the low-temperature
behavior as
\begin{equation}
   \frac{C}{Nk_{\rm B}}
   =\Bigl(
     \frac{k_{\rm B}T}{\pi aJ}
    \Bigr)^{1/2}
    \Bigl[
     \Bigl(
      \frac{\mit\Delta}{k_{\rm B}T}
     \Bigr)^2
    + \frac{\mit\Delta}{k_{\rm B}T}
    + \frac{3}{4}
    \Bigl]{\rm e}^{-{\mit\Delta}/k_{\rm B}T}.
\end{equation}
Considering the difficulty of grand canonical sampling at low temperatures
by numerical tools such as quantum Monte Carlo and density-matrix
renormalization group, the SFs can play an effective role in our
explorations.
Figure \ref{F:Th}(b) shows the magnetic susceptibility $\chi$.
The SFs overestimate the round peak at intermediate temperatures but again
work well at low temperatures revealing the initial
increase as
\begin{equation}
   \frac{\chi J}{N(g\mu_{\rm B})^2}
   =\Bigl(\frac{J}{\pi ak_{\rm B}T}\Bigr)^{1/2}
    {\rm e}^{-{\mit\Delta}/k_{\rm B}T}.
\end{equation}
\vspace*{-4mm}
\begin{figure}
\centerline
{\mbox{\psfig{figure=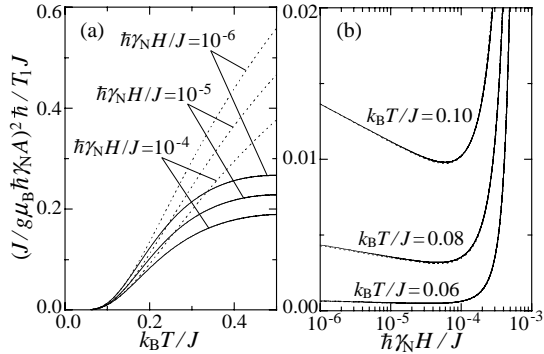,width=72mm,angle=0}}}
%\vspace*{1mm}
\caption{Temperature (a) and field (b) dependences of the nuclear
         spin-lattice relaxation rate calculated by the HFSFs.
         Numerical estimates of Eq. (\ref{E:T1SF}) (solid lines) are
         compared with the analytic expression (\ref{E:aT1SF})
         (dotted lines).}
\label{F:T1SF}
\end{figure}
\noindent
The MSWs are good at reproducing the overall behavior.
All the calculations converge into the paramagnetic susceptibility
$2S(S+1)/3k_{\rm B}T$ at high temperatures.

   Lastly we investigate the nuclear spin-lattice relaxation rate in terms
of the SFs, which are much better than the MSWs at describing the spin gap
and therefore the low-temperature properties.
Considering the electronic-nuclear energy-conservation requirement,
the Raman process plays a leading role in the relaxation, which is
formulated as
\begin{eqnarray}
   &&
   \frac{1}{T_1}
    =\frac{4\pi\hbar(g\mu_{\rm B}\gamma_{\rm N})^2}
          {\sum_n{\rm e}^{-E_n/k_{\rm B}T}}
     \sum_{n,m}{\rm e}^{-E_n/k_{\rm B}T}
   \nonumber\\
   &&\ \times
     \big|
      \langle m|AS_{i,j}^z|n\rangle
     \big|^2
     \,\delta(E_m-E_n-\hbar\omega_{\rm N}),
\label{E:T1def}
\end{eqnarray}
where $A$ is the hyperfine coupling constant between the nuclear and
electronic spins, $\omega_{\rm N}\equiv\gamma_{\rm N}H$ is the Larmor
frequency of the nuclei with $\gamma_{\rm N}$ being the gyromagnetic
ratio, and the summation $\sum_n$ is taken over all the electronic
eigenstates $|n\rangle$ with energy $E_n$.
Taking account of the significant difference between the electronic
and nuclear energy scales ($\hbar\omega_{\rm N}\alt 10^{-5}J$) and
assuming a reasonable temperature range
$k_{\rm B}T\alt{\mit\Delta}$ for SrCu$_2$O$_3$ with
${\mit\Delta}/k_{\rm B}\simeq 420\,\mbox{K}$ \cite{A3463}, the relaxation
rate is represented as
\begin{equation}
   \frac{1}{T_1}\simeq
    \frac{(g\mu_{\rm B}\hbar\gamma_{\rm N}A)^2}{4\pi\hbar Ja}
    \int_0^{2\pi}
    \frac{\sum_i\bar{n}_{i,k}(1-\bar{n}_{i,k})}
         {\sqrt{(k-\pi)^2+\hbar\omega_{\rm N}/Ja}}
    {\rm d}k,
    \label{E:T1SF}
\end{equation}
where we have again employed the approximate
dispersion (\ref{E:adsp}).
The term $\bar{n}_{i,k}(1-\bar{n}_{i,k})$ is the consequence of the
principle of detailed balancing in a fermionic ensemble.
At moderate fields and temperatures,
$k_{\rm B}T\ll{\mit\Delta}-g\mu_{\rm B}H$,
Eq. (\ref{E:T1SF}) can be further calculated analytically as
\begin{equation}
   \frac{1}{T_1}\simeq
    \frac{(g\mu_{\rm B}\hbar\gamma_{\rm N}A)^2}{2\pi\hbar Ja}
    {\rm e}^{-{\mit\Delta}/k_{\rm B}T}
    {\rm cosh}\frac{g\mu_{\rm B}H}{k_{\rm B}T}
    K_0\Bigl(\frac{\hbar\omega_{\rm N}}{2k_{\rm B}T}\Bigr),
   \label{E:aT1SF}
\end{equation}
where $K_0$ is the modified Bessel function of the second kind and behaves
as $K_0(x)\simeq 0.11593-{\rm ln}x$ for $0<x\ll 1$.
Equation (\ref{E:T1SF}), together with its approximate expression
(\ref{E:aT1SF}), is plotted in Fig. \ref{F:T1SF}.

   At low temperatures, $1/T_1$ also exhibits an increase of the
activation type but with logarithmic correction, which is much weaker than
the power correction in the case of the susceptibility.
Such a pure spin-gap-activated temperature dependence of $1/T_1$, which
was pointed out in a sophisticated numerical work \cite{T13515} as well,
may be observed at sufficiently low temperatures but is not yet verified
for ladder antiferromagnets such as SrCu$_2$O$_3$ owing to the magnetic
impurities masking the intrinsic properties.
Equation (\ref{E:T1SF}) deviates from the simple spin-gap-activated
behavior (\ref{E:aT1SF}) for $k_{\rm B}T\agt 0.2J\simeq 0.4{\mit\Delta}$,
which is interestingly consistent with the criterion for the breakdown of
the simple two-magnon scheme observed in a spin-$1$ Haldane-gap
antiferromagnet AgVP$_2$S$_6$ \cite{T2173}.
In such a high-temperature range, we have to fully incorporate particle
collisions into the calculation in order to describe the crossover to
diffusive behavior, where an enhanced activation gap may be observed due
to the temperature-dependent diffusion constant \cite{S943}.
In the low-temperature range $k_{\rm B}T\alt 0.2J$, on the other hand, 
the relaxation rate should strictly follow the expression (\ref{E:aT1SF}):
{\it With increasing field, $1/T_1$ first decreases logarithmically and
then increases exponentially}.
The initial logarithmic behavior comes from the Van Hove singularity
peculiar to one-dimensional energy spectra and may arise from a nonlinear
dispersion relation at the band bottom in more general.
Therefore, besides spin-gapped antiferromagnets, one-dimensional ferro-
and ferrimagnets may exhibit similar field dependence \cite{H054409}.
The present logarithmic field dependence at low temperatures should and
could be distinguished from the $1/\sqrt{H}$ or ${\rm ln}(1/H)$ dependence
of diffusion-dominated dynamics \cite{H965,A420} at high temperatures.
The following exponential increase originates in the spin-$1$ excited
state lowering in energy with increasing field.
If we consider the strong-correlation compound SrCu$_2$O$_3$ with
$J/k_{\rm B}\simeq 840\,\mbox{K}$ \cite{A3463}, the minimum is supposed to
appear at $H\simeq 20\,\mbox{T}$.
In the case of molecular-based ladder systems \cite{H5392,D1599,W5168}
with much smaller exchange interactions, $1/T_1$ should reach a minimum
at much smaller fields.
We expect that such intrinsic features of the low-energy dynamics of
ladder antiferromagnets will be found experimentally.

   We have demonstrated new schemes of investigating Heisenberg two-leg
ladder antiferromagnets.
Although both fermionic and bosonic approaches possess advantages of their
own, {\it we stress the SFs along the snake-like path as one of the best
languages for the present system.}
One of the fatal weak points in the MSW description is nonvanishing
specific heat at high temperatures.
The endlessly increasing energy with increasing temperature is because of
the temperature-dependent energy spectrum (\ref{E:dspMSW}), where
$\lambda$, playing the role of the chemical potential, turns out a
monotonically increasing function of temperature.
Such a difficulty can be overcome in ferrimagnets exhibiting
noncompensating sublattice magnetizations \cite{N214418,Y157603} but
plagues bosonic approaches applied to antiferromagnets.
The SFs well reproduce the spin gap and thus reliably describe the
low-temperature properties.
{\it One of the most interesting findings is the novel field dependence of
$1/T_1$.}
Relevant relaxation-time measurements are strongly encouraged.

   The authors are grateful to Professor M. Takigawa for useful comments.
This work was supported by the Ministry of Education, Culture, Sports,
Science and Technology of Japan and the Nissan Science Foundation.

\widetext
\end{document}